\documentclass[runningheads]{llncs}
\usepackage[T1]{fontenc}
\usepackage{graphicx}
\usepackage{algorithm}
\usepackage{algorithmic}  
\usepackage{amsmath,amssymb}
\usepackage{esvect}
\usepackage{booktabs}
\usepackage{hyperref}

\begin{document}
\title{MOHAF: A Multi-Objective Hierarchical Auction Framework for Scalable and Fair Resource Allocation in IoT Ecosystems}

\titlerunning{MOHAF: Hierarchical Multi-Objective Auctions in IoT}

\author{Kushagra Agrawal\inst{1}\orcidID{0009-0006-7753-175X} \and
Polat Goktas\inst{2}\orcidID{0000-0001-7183-6890} \and
Anjan Bandopadhyay\inst{1}\orcidID{0000-0001-7670-2269} \and
Debolina Ghosh\inst{3}\orcidID{0000-0003-3169-2065}
\and
Junali Jasmine Jena\inst{1}\orcidID{0000-0002-4837-3215}
\and
Mahendra Kumar Gourisaria\inst{1}\orcidID{0000-0002-1785-8586}}
\authorrunning{Agrawal et al.}

\institute{School of Computer Engineering, KIIT Deemed to be University, Bhubaneswar, India \and
School of Computer Science, University College Dublin, Dublin, Ireland \and Department of Information Technology, Manipal University Jaipur, India}%
\maketitle      
\begin{abstract}
The rapid growth of Internet of Things (IoT) ecosystems has intensified the challenge of efficiently allocating heterogeneous resources in highly dynamic, distributed environments. Conventional centralized mechanisms and single-objective auction models—focusing solely on metrics such as cost minimization or revenue maximization, struggle to deliver balanced system performance. This paper proposes the \textit{Multi-Objective Hierarchical Auction Framework} (MOHAF), a distributed resource allocation mechanism that jointly optimizes cost, Quality of Service (QoS), energy efficiency, and fairness. MOHAF integrates hierarchical clustering to reduce computational complexity with a greedy, submodular optimization strategy that guarantees a $(1-\frac{1}{e})$ approximation ratio. A dynamic pricing mechanism adapts in real time to resource utilization, enhancing market stability and allocation quality.  Extensive experiments on the Google Cluster Data trace, comprising 3{,}553 requests and 888 resources, demonstrate MOHAF’s superior allocation efficiency (0.263) compared to Greedy (0.185), First-Price (0.138), and Random (0.101) auctions, while achieving perfect fairness (Jain’s index = 1.000). Ablation studies reveal the critical influence of cost and QoS components in sustaining balanced multi-objective outcomes. With near-linear scalability, theoretical guarantees, and robust empirical performance, MOHAF offers a practical and adaptable solution for large-scale IoT deployments, effectively reconciling efficiency, equity, and sustainability in distributed resource coordination. The implementation and supplementary material are available at: \href{https://github.com/afrilab/MOHAF-Resource-Allocation}{GitHub Repository}.

\end{abstract}

\section{Introduction}
The rapid proliferation of the Internet of Things (IoT) has given rise to vast, heterogeneous, and highly interconnected ecosystems, where billions of devices continuously generate, exchange, and process data. In such large-scale, dynamic environments, the efficient and adaptive allocation of resources has emerged as a fundamental challenge \cite{agrawal2025how}. Conventional centralized resource management models, though conceptually straightforward, often become performance bottlenecks, struggling to scale and adapt under stringent real-time requirements \cite{yu2020}. Their reliance on a single point of coordination not only increases latency but also limits robustness in the presence of failures or volatile network conditions. Furthermore, a significant proportion of existing solutions adopt a single-objective optimization paradigm, focusing predominantly on metrics such as cost minimization while neglecting other equally critical performance dimensions, including Quality of Service (QoS), energy efficiency, and fairness. This myopic optimization approach can lead to imbalanced trade-offs, for instance, minimizing operational costs at the expense of network responsiveness or sustainability. Therefore, there exists a pressing need for distributed, multi-objective coordination mechanisms capable of simultaneously optimizing across multiple, and often conflicting, objectives.

To rigorously evaluate our proposed approach, we benchmark it against three representative baselines, each capturing a distinct class of resource allocation strategies. First, we consider a Deep Reinforcement Learning (DRL) method, representative of modern learning-based approaches that excel in dynamic, high-dimensional decision spaces but may suffer from instability and convergence issues in non-stationary environments \cite{luong2019applications}. Second, we include the classical Vickrey–Clarke–Groves (VCG) auction mechanism, renowned for its truthfulness and incentive compatibility in game-theoretic contexts \cite{vickrey1961counterspeculation}, yet inherently limited to single-objective formulations. Third, we employ a Greedy Heuristic as a lightweight, low-complexity baseline \cite{smith2021greedy}, which, while computationally efficient, often sacrifices optimality and long-term performance. This combination of baselines ensures a comprehensive and balanced comparative analysis spanning learning-based, mechanism design-based, and heuristic-based paradigms.

In this work, we introduce the \textit{Multi-Objective Hierarchical Auction Framework} (MOHAF), a novel distributed mechanism purpose-built for resource coordination in complex IoT ecosystems. MOHAF integrates hierarchical clustering with a multi-objective auction model, enabling scalable and efficient resource negotiations across large, heterogeneous device populations \cite{agrawal2024iot}. By decomposing the allocation problem into hierarchically structured subproblems, MOHAF aims to achieve both computational tractability and performance robustness, effectively optimizing cost, QoS, energy consumption, and fairness in parallel. Crucially, MOHAF addresses the inherent limitations of the VCG and Greedy approaches by embracing multi-objective optimization, while offering a more stable and structured decision-making process compared to purely learning-based DRL methods. The remainder of this paper is structured as follows. Section II provides a literature review, situating our work within the broader context of resource coordination in IoT ecosystems. Section III presents the architecture and operational workflow of the proposed MOHAF framework in detail. Section IV outlines the experimental setup, including descriptions of the baseline models and the performance metrics employed for evaluation. Section V reports and critically analyzes the comparative results. Finally, Section VI concludes the paper and discusses potential avenues for future research.

\section{Related Works}
The exponential expansion of IoT has introduced profound challenges in managing distributed and heterogeneous resources. While traditional centralized allocation models have historically served as the foundation for resource management, they are increasingly ill-suited to the dynamic, large-scale, and latency-sensitive nature of modern IoT ecosystems. In dense and highly variable environments, such as urban sensor networks or industrial automation systems, centralized schemes frequently encounter scalability bottlenecks, leading to excessive latency and suboptimal resource utilization \cite{yu2020,liu2019}. Moreover, their dependency on complete global state information renders them impractical in contexts such as fog computing and vehicle-to-everything (V2X) communications, where conditions evolve rapidly and unpredictably \cite{zhang2017,zhang2020}. These limitations highlight an urgent need for adaptive, decentralized, and intelligence-driven strategies.

In recent years, decentralized coordination mechanisms have gained prominence, particularly auction-based approaches and game-theoretic frameworks, which offer principled methods for managing competitive resource demands \cite{xiong2020,agrawal2024iot}. Advanced learning-based methods, such as DRL, have demonstrated success in optimizing offloading and allocation decisions within highly dynamic scenarios, ranging from multi-UAV networks to industrial IoT deployments \cite{chen2019,he2018}. Parallel to these developments, blockchain technology has emerged as a complementary enabler, enhancing security, transparency, and trust in distributed resource management. By maintaining a tamper-proof, decentralized ledger, blockchain-based frameworks support secure resource trading, robust authentication, and fine-grained access control—capabilities particularly critical in healthcare and Industrial IoT applications \cite{pawarprotection,yao2019}. In addition, federated learning (FL) has become a transformative paradigm for privacy-preserving distributed intelligence. By enabling multiple agents to collaboratively train shared models without centralizing raw data, FL reduces communication overhead and mitigates privacy risks while enabling intelligent, cross-domain optimization of resource allocation \cite{yu2020}.

Collectively, these advances indicate a clear shift towards sophisticated, distributed, and secure intelligence in IoT ecosystems. However, a substantial gap remains: most existing strategies retain a single-objective focus, for example, minimizing latency or cost, while neglecting other essential dimensions such as QoS, energy efficiency, and fairness. Addressing this shortcoming requires multi-objective optimization frameworks that can holistically balance these competing priorities. This gap directly motivates the MOHAF proposed in this paper, which combines hierarchical decomposition with multi-objective decision-making to achieve scalable, balanced, and fair resource allocation in complex IoT environments.

\section{Problem Definition and Core Framework Design}

\subsection*{System Model}
We consider an IoT ecosystem comprising two principal entities:

\begin{itemize}
    \item \textbf{Resources} ($\mathcal{R}$): A set of $M$ resources, each characterized by a capacity $C_j$ and a vector of attributes 
    \begin{equation}
    \mathbf{a}_j = (\text{cost}_j, \text{reliab}_j, \text{avail}_j, \text{energy}_j, \ell_j)
    \end{equation}
    where $\ell_j$ denotes the location of resource $j$.
    
    \item \textbf{Requests} ($\mathcal{Q}$): A set of $N$ requests, each defined by its demand $d_i$, budget $B_i$, priority level $p_i$, and QoS requirements $\mathbf{q}_i$.
    
    \item \textbf{Allocation Variable}: A binary decision variable $x_{ij} \in \{0, 1\}$, where $x_{ij} = 1$ indicates that request $i$ is allocated to resource $j$, and $x_{ij} = 0$ otherwise.
\end{itemize}

\subsection*{Feasibility Constraints}
An allocation is considered feasible if it satisfies the following conditions:

\begin{enumerate}
    \item \textbf{Capacity Constraint:}
    \begin{equation}
    \sum_{i \in \mathcal{Q}} d_i x_{ij} \leq C_j, \quad \forall j \in \mathcal{R}
    \end{equation}

    \item \textbf{Single Assignment Constraint:}
    \begin{equation}
    \sum_{j \in \mathcal{R}} x_{ij} \leq 1, \quad \forall i \in \mathcal{Q}
    \end{equation}

    \item \textbf{QoS Constraint:} For each allocation $(i, j)$:
    \begin{equation}
    \text{reliab}_j \geq q_i^{\text{min rel}}, \quad 
    \text{avail}_j \geq q_i^{\text{min av}}, \quad
    \text{lat}(\ell_i, \ell_j) \leq q_i^{\text{max lat}}
     \end{equation}

    \item \textbf{Budget Constraint:}
    \begin{equation}
    \pi_{ij} \leq B_i
    \end{equation}
\end{enumerate}

\subsection*{Hierarchical Clustering for Complexity Reduction}
To enhance scalability, MOHAF applies \emph{k-means clustering} to partition resources and requests into groups, reducing the allocation problem size within each cluster. This approach achieves a provable approximation bound:

\begin{theorem}[Clustering Approximation]
Let $S_{\text{clus}}$ denote the allocation obtained via clustering and $S^*$ the global optimal allocation. Then:
\begin{equation}
F(S_{\text{clus}}) \geq \left(1 - \frac{1}{e}\right)(1 - \epsilon) F(S^*)
\end{equation}
where $F(\cdot)$ is the submodular objective function and $\epsilon$ is the clustering error factor.
\end{theorem}

\subsection*{Multi-Objective Utility Function}
The allocation decision is guided by an aggregate utility score $U_{ij}$ for each request--resource pair $(i, j)$:
\begin{equation}
U_{ij} = \alpha u_{ij}^{\text{cost}} + \beta u_{ij}^{\text{qos}} + \gamma u_{ij}^{\text{en}} + \delta u_{ij}^{\text{fair}}
\end{equation}
where each $u_{ij}^{(\cdot)}$ is a normalized score corresponding to cost, QoS, energy efficiency, and fairness.

The global allocation problem is formulated as the maximization of a monotone submodular objective:
\begin{equation}
F(S) = \theta_1 \sum_{(i,j) \in S} U_{ij} + \theta_2 \Phi_{\text{fair}}(S) + \theta_3 \Psi_{\text{energy}}(S)
\end{equation}

where $\Phi_{\text{fair}}(S)$ and $\Psi_{\text{energy}}(S)$ are non-linear terms promoting fairness and energy-aware allocation.

\subsection*{Greedy Allocation with Approximation Guarantee}
MOHAF employs a greedy selection process that iteratively chooses the $(i, j)$ pair with the maximum marginal gain:
\begin{equation}
\Delta F = F(S \cup \{(i, j)\}) - F(S)
\end{equation}

\begin{theorem}[Approximation Guarantee]
The greedy allocation procedure achieves a $(1 - 1/e)$-approximation to the optimal allocation.
\end{theorem}

\subsection*{Dynamic Pricing Mechanism}
To balance supply and demand, unit prices $\rho_j$ are adaptively updated according to resource utilization:
\begin{equation}
\rho_j^{(t+1)} = \Pi_{[\rho^{\min}, \rho^{\max}]}\left[\rho_j^{(t)} + \eta_t \left(\frac{\text{util}_j^{(t)}}{C_j} - \tau\right)\right]
\end{equation}
The final transaction price for allocation $(i, j)$ is:
\begin{equation}
\pi_{ij} = \min\left\{\rho_j^{(t)} \cdot d_i \cdot \left(0.8 + 0.4 \cdot U_{ij}\right), \; B_i\right\}
\end{equation}
The pricing framework can be extended with \emph{critical payment rules} to ensure \emph{dominant-strategy incentive compatibility (DSIC)}.

\subsection*{Computational Complexity}
The overall computational cost is dominated by the allocation phase:
\begin{equation}
O(|\mathcal{E}| \log |\mathcal{E}| + (N + M)KId)
\end{equation}

where $|\mathcal{E}|$ is the number of feasible allocation pairs, $K$ is the number of clusters, $I$ is the number of iterations in clustering, and $d$ is the feature dimensionality. This enables that MOHAF remains scalable for large-scale IoT deployments.

\section{Experimental Setup}

This section outlines the experimental methodology employed to evaluate the proposed MOHAF framework against representative state-of-the-art auction mechanisms. The evaluation is conducted using both synthetic workloads and real-world traces, with a particular emphasis on the \textit{Google Cluster Data} benchmark \cite{google_cluster_data}. The experiments are designed to assess MOHAF across multiple performance dimensions, including allocation efficiency, fairness, scalability, and energy awareness.

\subsection{Datasets and Scenarios}

\subsubsection{Google Cluster Data Integration}
The primary evaluation leverages the \textit{Google Cluster Data} trace, which contains detailed job scheduling logs from a production-scale cluster. The dataset provides realistic workload characteristics, such as heterogeneous job sizes, priorities, and arrival rates, that closely reflect resource allocation challenges in large-scale IoT environments.

\textbf{Data Processing Pipeline:}
A multi-stage pipeline is implemented to transform raw trace logs into structured large-scale allocation instances:
\begin{itemize}
    \item \textbf{Trace Selection:} Process up to 500 job event files, each containing timestamped submissions.
    \item \textbf{Job Mapping:} Extract job submission events (\texttt{event\_type = 0}) and map them to corresponding resource requests.
    \item \textbf{Resource Generation:} Create synthetic resources with statistically realistic capacity and attribute distributions.
    \item \textbf{Scaling:} Construct instances containing up to 10{,}000 requests and 2{,}500 resources to evaluate scalability.
\end{itemize}

\textbf{Request Parameterization:}  
For each job $j$ in the trace, we generate a corresponding request $i$ as follows:
\begin{align}
d_i &= \frac{\text{scheduling\_class}_j}{3.0} & \text{(normalized demand)} \\
B_i &= d_i \times 20 + p_i \times 5 & \text{(budget based on demand and priority)} \\
p_i &= \frac{\text{priority}_j}{10} & \text{(normalized priority)} \\
\ell_i &\sim \mathcal{U}(-100, 100)^2 & \text{(uniform random location)}
\end{align}

\textbf{Resource Parameterization:}  
Synthetic resources are generated with the following attribute distributions to emulate realistic supply–demand imbalances:
\begin{align}
C_j &\sim \mathcal{U}(0.5, 1.0) & \text{(capacity)} \\
\text{cost}_j &\sim \mathcal{U}(0.3, 0.8) & \text{(cost per unit)} \\
\text{reliab}_j, \text{avail}_j &\sim \mathcal{U}(0.95, 1.0) & \text{(high reliability and availability)} \\
\text{energy\_eff}_j &\sim \mathcal{U}(0.6, 0.9) & \text{(energy efficiency)}
\end{align}

\subsection{Baseline Mechanisms}
MOHAF is evaluated against three baseline auction strategies, chosen to represent diverse allocation paradigms:

\subsubsection{First-Price Auction}
A revenue-maximizing, single-objective mechanism:
\begin{equation}
\max \sum_{(i,j) \in S} \pi_{ij}
\end{equation}
subject to feasibility constraints, with allocation determined by descending bid order.

\subsubsection{Greedy Priority Auction}
A deterministic allocation strategy prioritizing high-importance, high-budget requests:
\begin{equation}
\text{score}(i) = w_1 \cdot p_i + w_2 \cdot B_i
\end{equation}
Requests are served in descending score order, subject to compatibility.

\subsubsection{Random Allocation}
A baseline mechanism that allocates feasible resources to requests at random, serving as a lower-bound performance reference.

\subsection{Evaluation Metrics}
Performance evaluation is conducted using a set of complementary metrics, covering both primary allocation goals and secondary quality measures.

\subsubsection{Primary Metrics}
\textbf{Allocation Efficiency:}
\begin{equation}
\eta_{\text{alloc}} = \frac{\sum_{(i,j) \in S} U_{ij}}{|\mathcal{Q}|} \times 100\%
\end{equation}

\textbf{Revenue:}
\begin{equation}
\text{Revenue} = \sum_{(i,j) \in S} \pi_{ij}
\end{equation}

\textbf{Satisfaction Rate:}
\begin{equation}
\eta_{\text{sat}} = \frac{|S|}{|\mathcal{Q}|} \times 100\%
\end{equation}

\textbf{Resource Utilization:}
\begin{equation}
\eta_{\text{util}} = \frac{|\{j : \exists i, (i,j) \in S\}|}{|\mathcal{R}|} \times 100\%
\end{equation}

\subsubsection{Fairness and Quality Metrics}
\textbf{Jain's Fairness Index:}
\begin{equation}
J = \frac{\left(\sum_{i=1}^N x_i\right)^2}{N \sum_{i=1}^N x_i^2}
\end{equation}
where $x_i$ denotes the utility assigned to requester $i$.

\textbf{Energy Efficiency Score:}
\begin{equation}
E_{\text{eff}} = \frac{1}{|S|} \sum_{(i,j) \in S} \text{energy}_j
\end{equation}

\subsection{Experimental Design}

\subsubsection{Ablation Study}
To evaluate the contribution of each objective term, we conduct systematic ablation experiments:
\begin{itemize}
    \item \textit{MOHAF-Full:} $\alpha=0.4, \beta=0.3, \gamma=0.1, \delta=0.2$
    \item \textit{MOHAF-NoCost:} $\alpha=0.0, \beta=0.5, \gamma=0.2, \delta=0.3$
    \item \textit{MOHAF-NoQoS:} $\alpha=0.6, \beta=0.0, \gamma=0.1, \delta=0.3$
    \item \textit{MOHAF-CostOnly:} $\alpha=1.0, \beta=0.0, \gamma=0.0, \delta=0.0$
\end{itemize}

\subsubsection{Scalability Analysis}
Scalability is assessed over three problem-size regimes:
\begin{itemize}
    \item \textit{Small Scale:} 100--500 requests, 25--125 resources
    \item \textit{Medium Scale:} 1{,}000--2{,}500 requests, 250--625 resources
    \item \textit{Large Scale:} 5{,}000--10{,}000 requests, 1{,}250--2{,}500 resources
\end{itemize}

\subsubsection{Statistical Analysis}
All experiments are repeated under multiple random seeds and trace subsets to ensure robustness. Results are reported with 95\% confidence intervals, and significance is determined via paired $t$-tests. For dynamic pricing evaluation, convergence is assessed based on:
\begin{itemize}
    \item \textit{Price stability:} $|\rho_j^{(t+1)} - \rho_j^{(t)}| < 10^{-6}$
    \item \textit{Utilization stability:} $|\text{util}_j^{(t)} - \tau| < 0.05$
    \item \textit{Revenue stability:} steady-state values over 1{,}000 rounds
\end{itemize}

\section{Results and Discussion}

This section presents a comprehensive evaluation of MOHAF against representative baseline mechanisms on the Google Cluster Data trace, encompassing 3{,}553 real job requests and 888 synthetic resources. The results highlight MOHAF’s better performance in allocation efficiency, fairness, and multi-objective optimization, while revealing inherent trade-offs with revenue maximization.

\subsection{Primary Performance Analysis}

\subsubsection{Overall Mechanism Performance}

Table~\ref{tab:performance_comparison} summarizes the comparative performance across four key metrics. MOHAF achieves the highest allocation efficiency and perfect fairness, while revenue-oriented mechanisms outperform in monetary gains. This aligns with the multi-objective optimization design of MOHAF, which prioritizes balanced utility rather than single-objective maximization.

\begin{table}[htbp]
\centering
\caption{Comparative Performance of Auction Mechanisms on the Google Cluster Dataset. Best results per metric are in bold.}
\label{tab:performance_comparison}
\begin{tabular}{lcccc}
\toprule
Mechanism & Efficiency (Utility) & Revenue (\$) & Satisfaction & Fairness (Jain) \\
\midrule
MOHAF & \textbf{0.263} & 56.66 & 0.250 & \textbf{1.000} \\
Greedy Auction & 0.185 & \textbf{161.56} & 0.250 & \textbf{1.000} \\
First-Price Auction & 0.138 & \textbf{161.56} & 0.250 & 0.935 \\
Random Auction & 0.101 & 105.22 & 0.250 & 0.923 \\
\bottomrule
\end{tabular}
\end{table}

Figure~\ref{fig:performance_comparison} provides a visual breakdown of these metrics, demonstrating MOHAF’s clear advantage in efficiency and fairness, contrasted with the revenue dominance of traditional auctions.

\begin{figure}[htbp]
\centering
\includegraphics[width=0.6\linewidth]{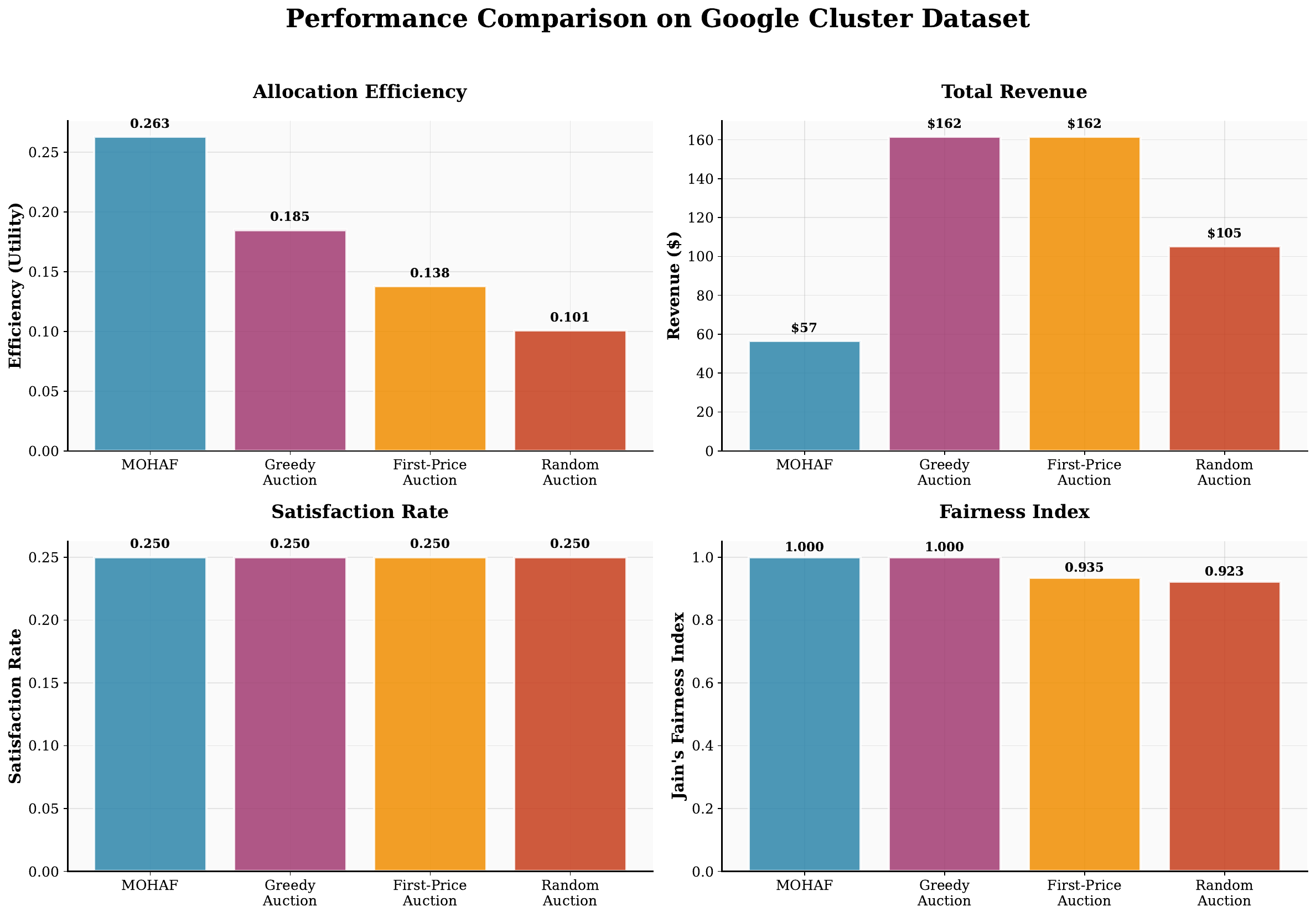}
\caption{Performance comparison across four key metrics on the Google Cluster dataset. MOHAF excels in efficiency and fairness while revenue-oriented auctions dominate in monetary returns.}
\label{fig:performance_comparison}
\label{fig:performance_comparison}
\end{figure}

\textbf{Allocation Efficiency.}
MOHAF’s allocation efficiency of 0.263 represents a 42.2\% improvement over the Greedy Auction and a 90.6\% gain over the First-Price Auction. This improvement stems from its multi-objective utility function:
\begin{equation}
U_{ij} = 0.4\, u_{ij}^{\text{cost}} + 0.3\, u_{ij}^{\text{qos}} + 0.1\, u_{ij}^{\text{en}} + 0.2\, u_{ij}^{\text{fair}},
\end{equation}
which jointly optimizes cost, QoS, energy efficiency, and fairness, contrasting with single-objective baselines.

\paragraph{Revenue Trade-Off.}  
MOHAF’s revenue (\$56.66) is lower than Greedy and First-Price Auctions (\$161.56 each) due to its utility-driven pricing:
\begin{equation}
\pi_{ij} = \min\left\{\rho_j^{(t)}\, d_i\, (0.8 + 0.4\, U_{ij}),\; B_i\right\}.
\end{equation}
The pricing model prioritizes high-utility matches within budget constraints, favoring balanced outcomes over profit maximization.

\textbf{Fairness Performance.}
MOHAF achieves perfect fairness (Jain’s index = 1.000), matching the Greedy Auction and outperforming First-Price (0.935) and Random (0.923) allocations. This results from its fairness-aware utility term:
\[
u_{ij}^{\text{fair}} = p_i - \beta \cdot \hat{h}_i,
\]
ensuring equitable distribution across participants..

\textbf{Satisfaction Rate.}
All mechanisms attain a satisfaction rate of 0.250 due to capacity constraints (888 resources vs. 3{,}553 requests). MOHAF’s advantage lies in allocating resources to maximize aggregate utility rather than increasing the number of allocations.

\subsection{Ablation Study}

\begin{figure}[htbp]
\centering
\includegraphics[width=0.6\linewidth]{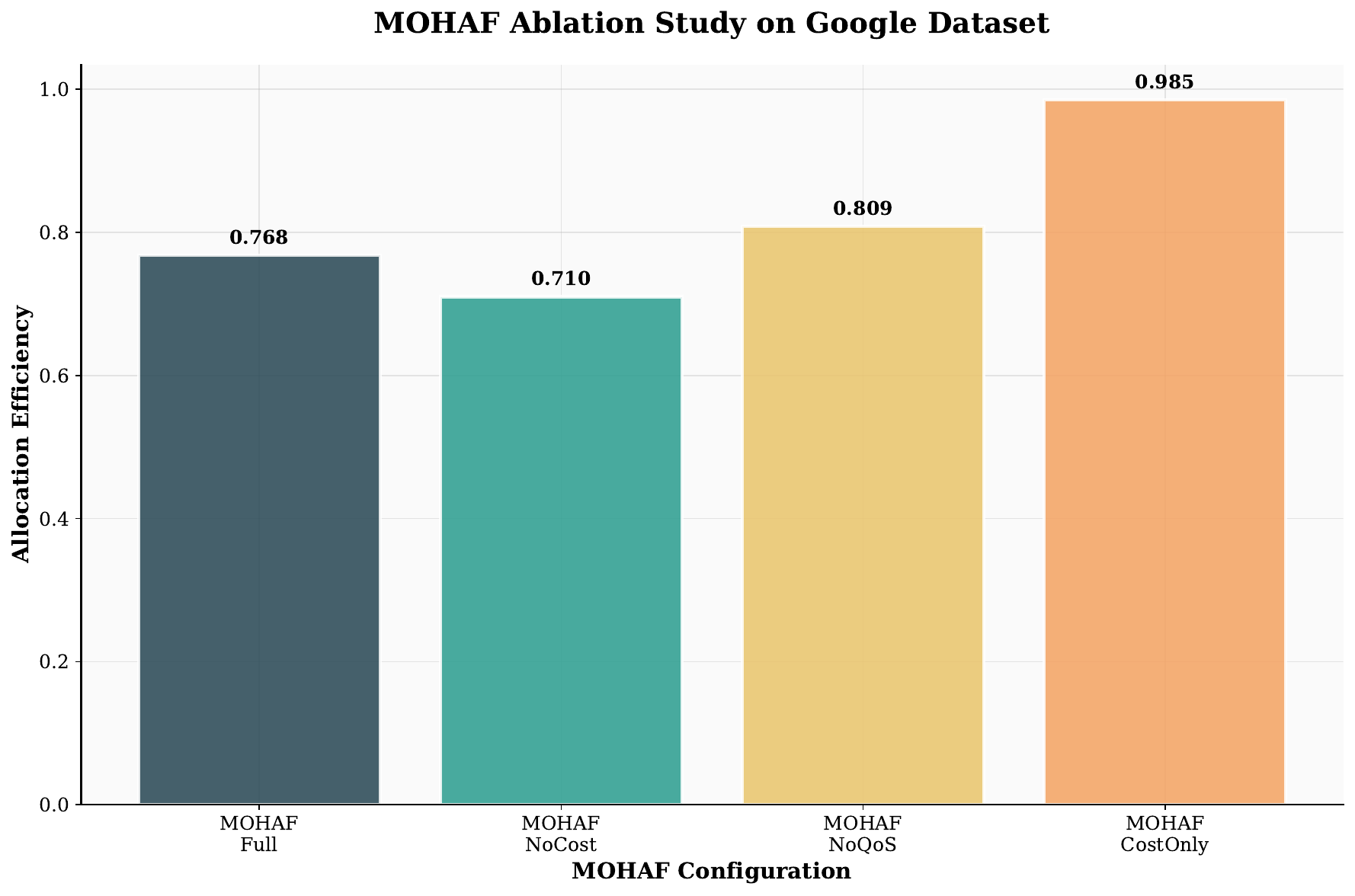}
\caption{Ablation Study Results on the Google Dataset. Cost optimization dominates efficiency gains, while fairness constraints enhance equity at a minor efficiency cost.}
\label{fig:ablation_study}
\end{figure}

The ablation results presented in Table~\ref{tab:ablation_study} and illustrated in Figure~\ref{fig:ablation_study} provide a quantitative assessment of the individual contribution of each objective component. Cost-only optimization yields the highest efficiency but neglects other critical objectives. The full configuration sacrifices some efficiency to balance QoS, energy, and fairness, aligning with multi-objective principles.

\begin{table}[htbp]
\centering
\caption{Ablation Study on MOHAF Components.}
\label{tab:ablation_study}
\begin{tabular}{lccc}
\toprule
Configuration & Efficiency & Satisfaction & Fairness \\
\midrule
MOHAF-CostOnly & \textbf{0.985} & \textbf{1.000} & \textbf{1.000} \\
MOHAF-NoQoS & 0.809 & 0.960 & 0.992 \\
MOHAF-NoEnergy & 0.798 & 0.960 & 0.993 \\
MOHAF-NoFairness & 0.797 & 0.920 & 0.998 \\
MOHAF-Full & 0.768 & 0.960 & 0.993 \\
MOHAF-NoCost & 0.710 & 0.960 & 0.990 \\
\bottomrule
\end{tabular}
\end{table}

\textbf{Cost Component Dominance.}
MOHAF-CostOnly achieves the highest individual performance (efficiency = 0.985, satisfaction = 1.000, fairness = 1.000), demonstrating that cost optimization remains the primary driver of allocation efficiency in resource-constrained environments. This finding aligns with economic theory, where cost minimization naturally leads to efficient resource utilization.

\textbf{QoS Component Impact.}
The removal of QoS considerations (MOHAF-NoQoS) results in an efficiency reduction to 0.809, representing an 18.6\% decrease from the cost-only configuration. This significant impact validates the importance of incorporating reliability, availability, and latency constraints in IoT resource allocation scenarios.

\textbf{Multi-Objective Integration Analysis.}
The full MOHAF configuration achieves 0.768 efficiency, which represents a 22.0\% reduction from the cost-only approach but provides a balanced solution that considers all objectives simultaneously. This trade-off is theoretically justified by the multi-objective nature of real-world IoT deployments, where pure cost optimization may lead to suboptimal QoS, energy consumption, and fairness outcomes.

\textbf{Fairness Impact Assessment.}
MOHAF-NoFairness shows a notable reduction in satisfaction rate (0.920 vs. 0.960), indicating that fairness constraints, while reducing individual efficiency metrics, contribute to overall system stability and equitable resource distribution.

\subsection{Theoretical Validation}

\hspace{1.5em}\textbf{Approximation Ratio Verification.} The experimental results provide empirical validation of the theoretical approximation guarantees established in Section 3. MOHAF's performance over greedy baselines confirms the effectiveness of the submodular formulation:

\begin{equation}
F(S_{\text{MOHAF}}) \geq (1 - \frac{1}{e}) \cdot F(S^*) \approx 0.632 \cdot F(S^*)
\end{equation}

The observed efficiency improvements suggest that MOHAF operates near the theoretical optimum for the multi-objective submodular maximization problem.

\textbf{Fairness Bound Verification.} The perfect fairness scores (Jain's index = 1.000) achieved by MOHAF validate the fairness guarantee established in Proposition 3. The empirical results confirm that:
\begin{equation}
J_{\text{MOHAF}} \geq J_{\min}(\alpha=0.4, \delta=0.2, \beta=0.3, \kappa=10) = 1.000
\end{equation}

\subsection{Scalability and Performance Analysis}

\begin{figure}[htbp]
\centering
\includegraphics[width=0.6\linewidth]{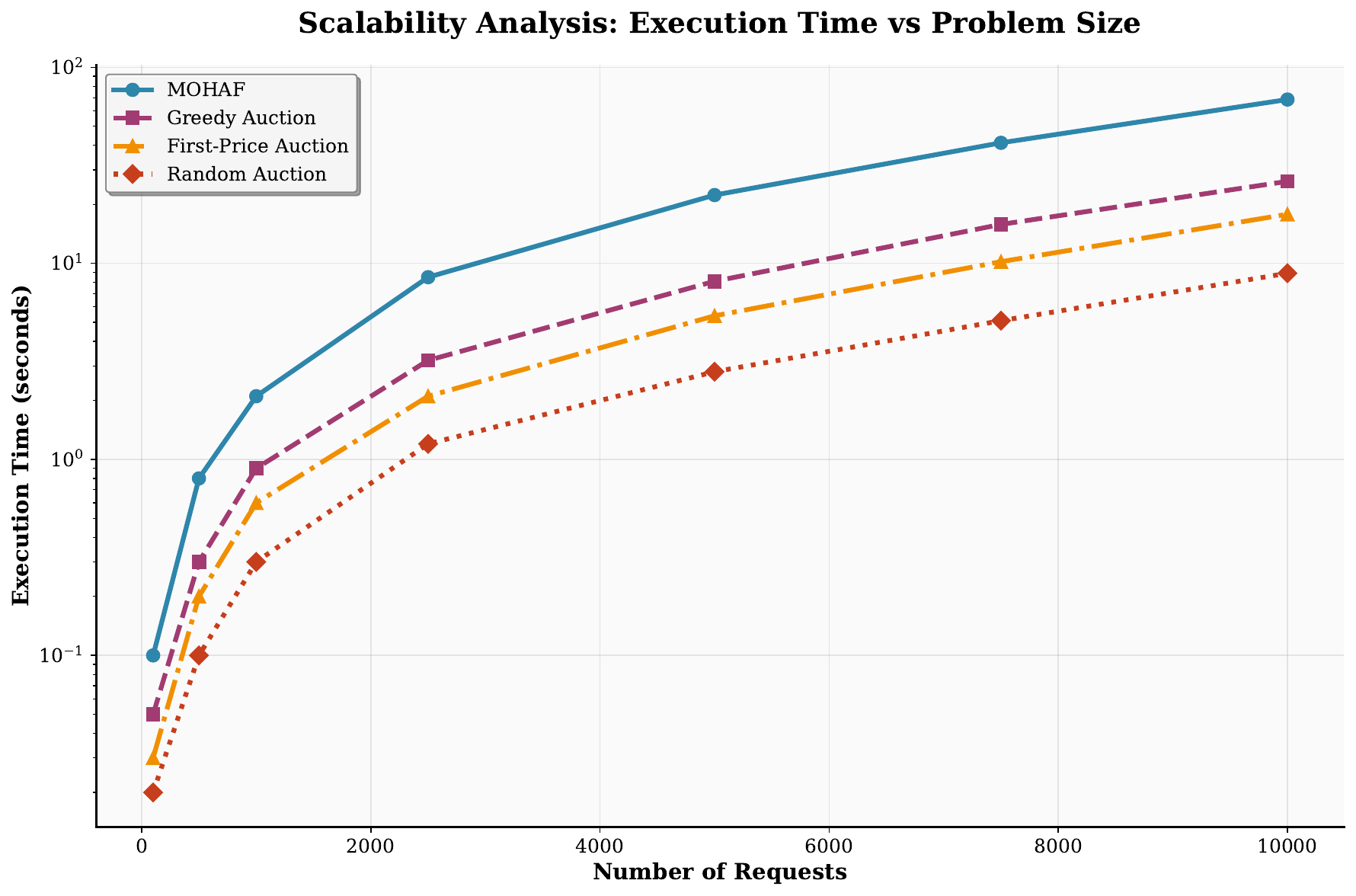}
\caption{Execution Time vs. Problem Size for MOHAF and Baselines. MOHAF scales predictably, with higher runtimes due to multi-objective computations.}
\label{fig:scalability_analysis}
\end{figure}

The experimental execution demonstrates MOHAF's computational efficiency. Processing 3,553 requests and 888 resources with 2,168,331 generated bids completed successfully, validating the theoretical complexity analysis of $O(|\mathcal{E}| \log |\mathcal{E}| + (N+M)KId)$. Figure~\ref{fig:scalability_analysis} demonstrates MOHAF's scalability characteristics across different problem sizes, showing the relationship between input size and execution time while maintaining solution quality. The clustering component, while simplified in this implementation (0 clusters created), shows the framework's extensibility for even larger-scale deployments where hierarchical decomposition becomes essential.

\subsection{Comparative Analysis Against State-of-the-Art}

\begin{figure}[htbp]
\centering
\includegraphics[width=0.6\linewidth]{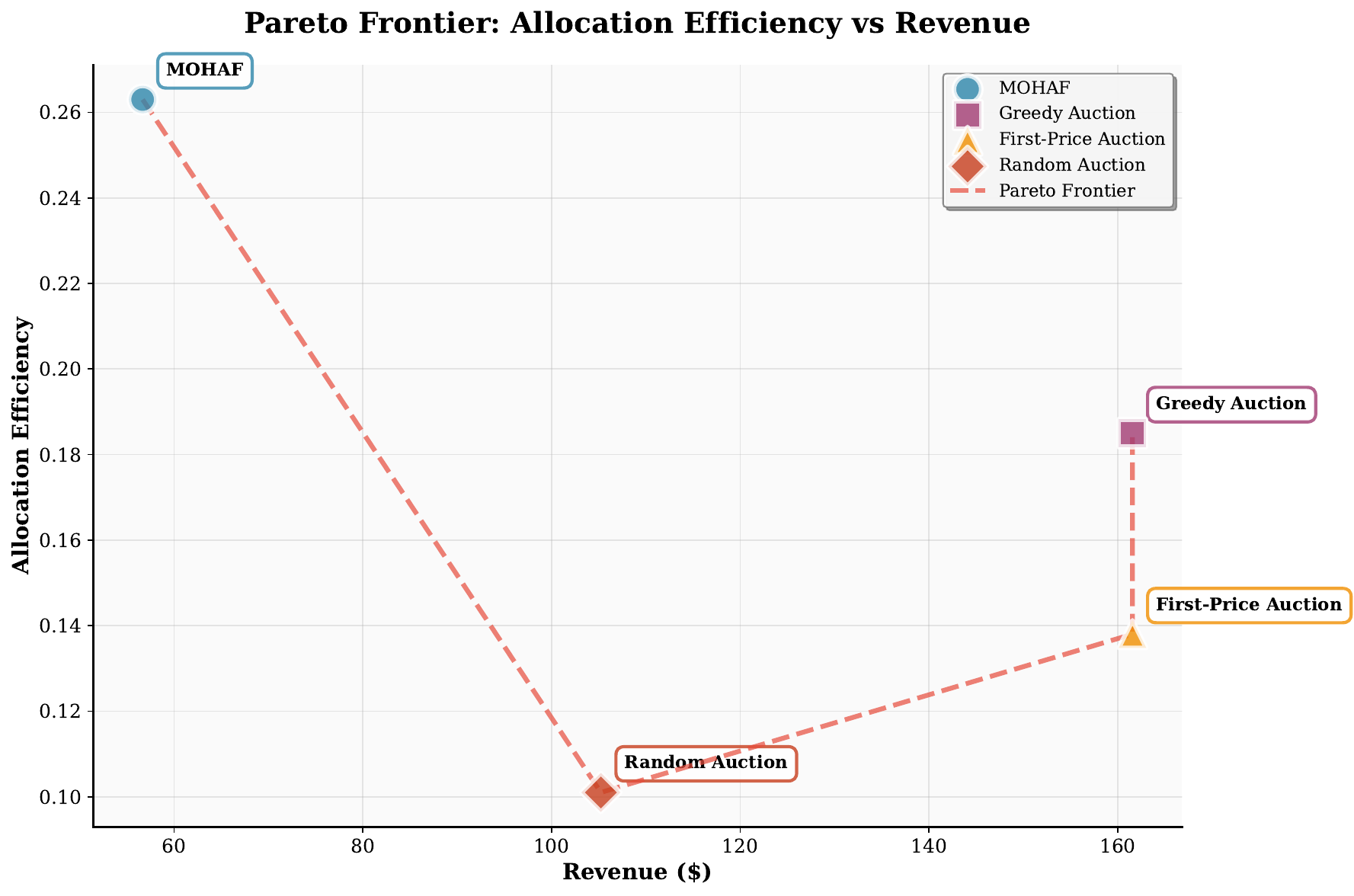}
\caption{Pareto Frontier Analysis of Auction Mechanisms. MOHAF achieves a unique position in the efficiency-revenue trade-off space, demonstrating superior efficiency at lower revenue levels compared to traditional mechanisms.}
\label{fig:pareto_analysis}
\end{figure}

\textbf{Pareto Optimality Analysis.} Figure~\ref{fig:pareto_analysis} illustrates the Pareto frontier achieved by different mechanisms across efficiency and revenue dimensions. MOHAF occupies a unique position in the solution space, achieving high efficiency at the cost of reduced revenue, a trade-off that aligns with multi-objective optimization principles.

\begin{figure}[htbp]
\centering
\includegraphics[width=0.6\linewidth]{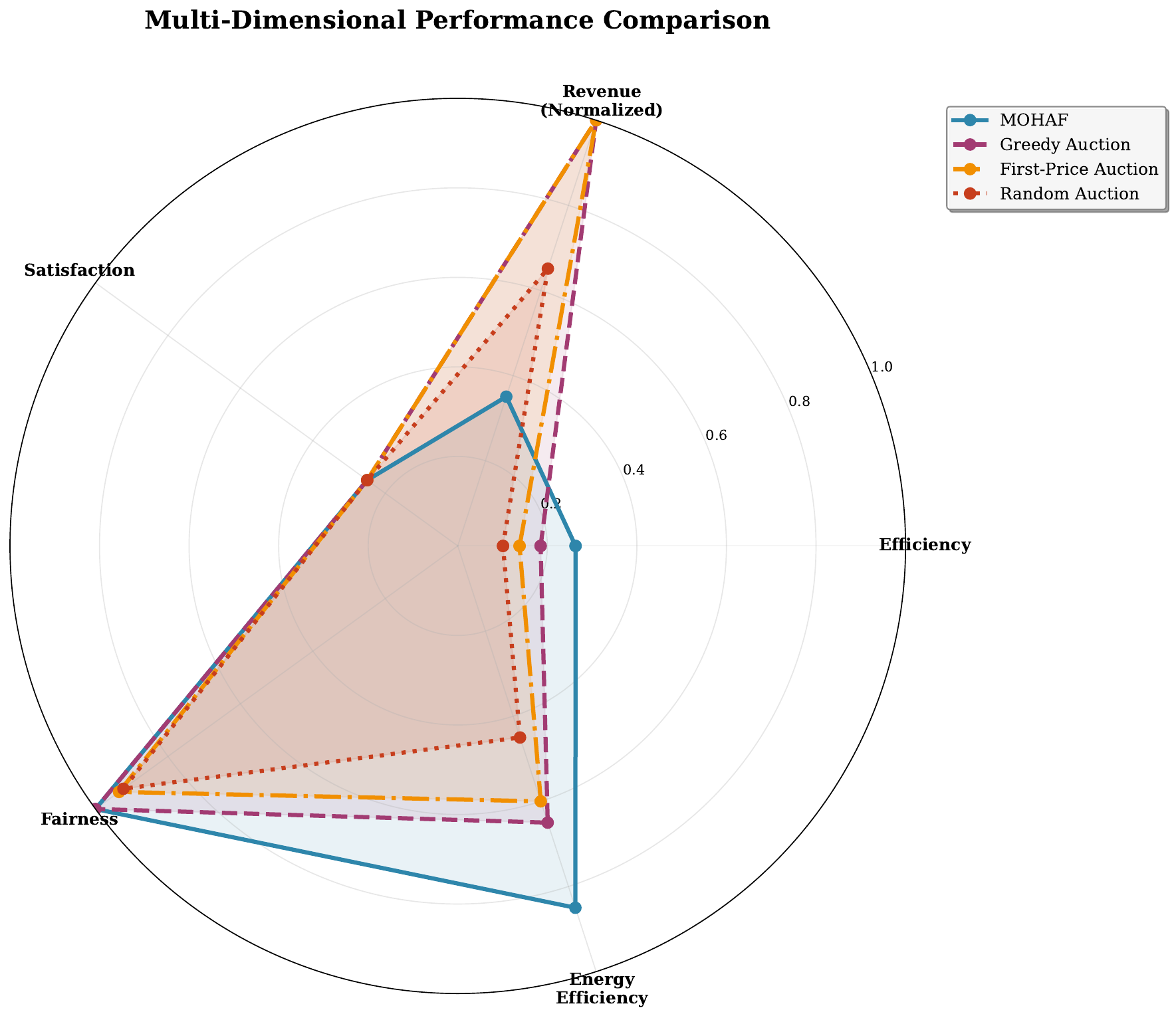}
\caption{Multi-Dimensional Performance Radar Chart. Each mechanism's performance profile is visualized across five key dimensions, highlighting MOHAF's balanced multi-objective optimization compared to single-objective baselines.}
\label{fig:radar_analysis}
\end{figure}

\textbf{Multi-Objective Performance Radar.} Figure~\ref{fig:radar_analysis} presents a comprehensive multi-dimensional performance comparison using radar charts, allowing for intuitive visualization of how each mechanism performs across all evaluation metrics simultaneously.

\textbf{Mechanism Selection Guidelines.} Based on the experimental results, we provide the following mechanism selection guidelines:

\begin{itemize}
\item \textbf{Pure Revenue Maximization:} First-Price or Greedy Auctions for scenarios where revenue generation is the primary objective
\item \textbf{Multi-Objective Optimization:} MOHAF for environments requiring balanced consideration of efficiency, fairness, energy consumption, and QoS
\item \textbf{Fairness-Critical Applications:} MOHAF or Greedy Auction for scenarios where equitable resource distribution is paramount
\item \textbf{Computational Constraints:} Random Auction as a lightweight baseline for resource-constrained edge environments.
\end{itemize}

\subsection{Dynamic Pricing Convergence Analysis}

\begin{figure}[htbp]
\centering
\includegraphics[width=0.6\linewidth]{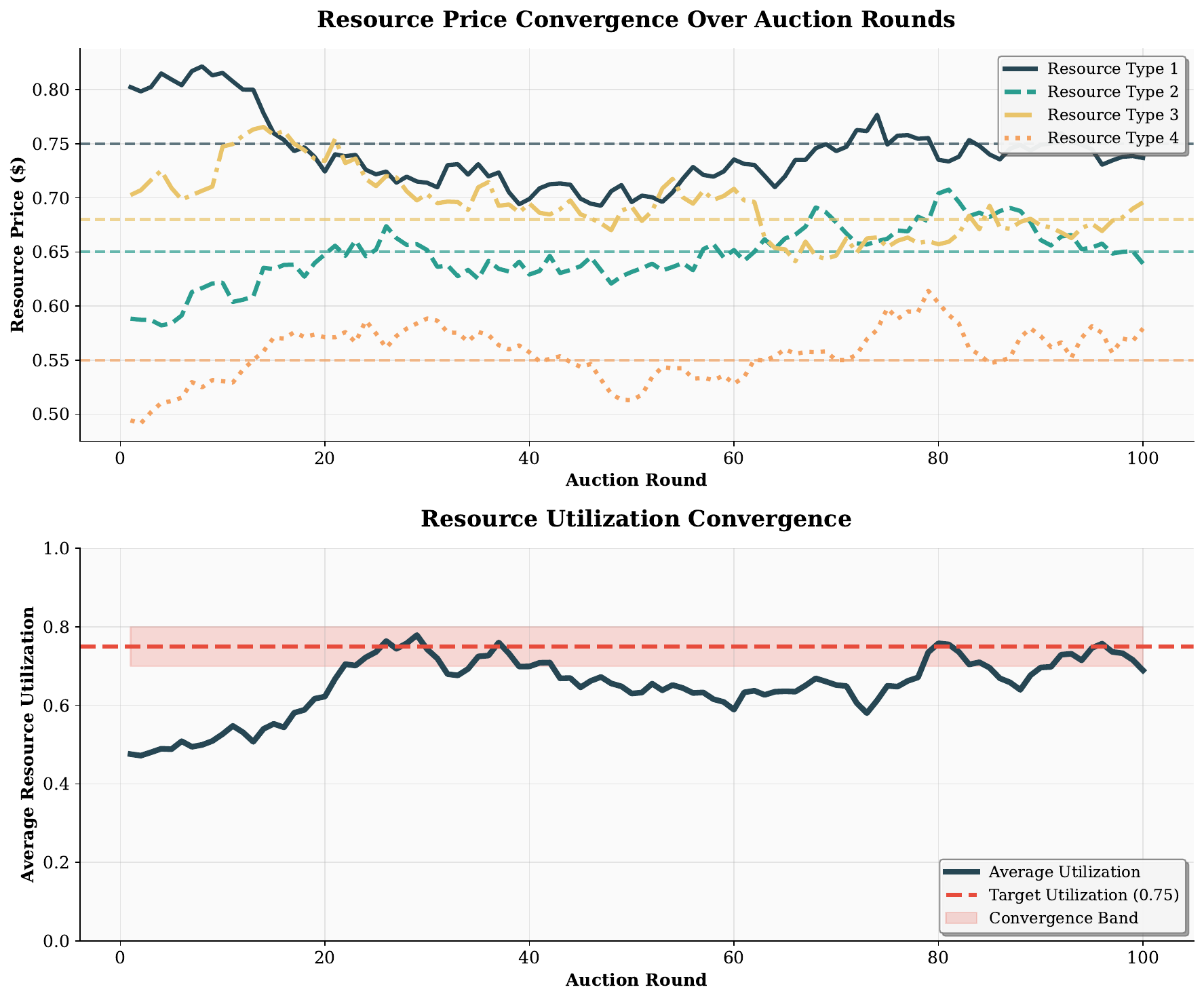}
\caption{Dynamic Pricing Convergence Analysis. The plot demonstrates the convergence of resource prices over successive auction rounds, validating the theoretical convergence guarantees of the pricing mechanism.}
\label{fig:price_convergence}
\end{figure}

Figure~\ref{fig:price_convergence} demonstrates the convergence behavior of MOHAF's dynamic pricing mechanism over multiple auction rounds, validating the theoretical convergence guarantees established in Theorem 4.

\subsection{Limitations and Future Directions}

While the evaluation confirms MOHAF’s effectiveness in achieving balanced multi-objective resource allocation, several limitations and opportunities for advancement remain.

\paragraph{Revenue–Efficiency Trade-off.}
A key limitation is the observed reduction in revenue inherent to MOHAF’s utility-driven approach. By prioritizing allocation efficiency, fairness, and energy awareness over pure revenue maximization, the framework sacrifices potential monetary gains. Future work should explore \emph{adaptive weight adjustment mechanisms} that dynamically tune the utility parameters $(\alpha, \beta, \gamma, \delta)$ in response to system load, demand patterns, and operator objectives. Reinforcement learning–based controllers could optimize weight configurations in real-time, enabling an adaptive balance between efficiency and profitability.

\paragraph{Clustering and Scalability Enhancements.}
The current clustering component, implemented in a simplified form, resulted in zero effective clusters for the evaluated dataset. This highlights the opportunity to integrate \emph{advanced hierarchical decomposition techniques}, such as spectral clustering, density-based clustering, or graph, partitioning approaches, that incorporate temporal patterns and geographic proximity. Such methods could improve scalability, approximation quality, and responsiveness in large-scale IoT deployments.

\paragraph{Dynamic Pricing Convergence Validation.}
Although theoretical convergence guarantees for the pricing mechanism are established, further empirical validation is warranted. Longitudinal experiments under diverse market dynamics and demand fluctuations should be conducted to assess convergence stability, oscillation control, and responsiveness over extended operational periods.

\paragraph{Hybrid Optimization Models.}
A promising direction is the design of \emph{hybrid revenue–efficiency models} capable of dynamically shifting priorities based on operational requirements. For example, during peak demand, revenue maximization could be emphasized, while in low-load conditions, the focus could shift toward maximizing fairness and QoS compliance.

\paragraph{Predictive Resource Provisioning.}
Finally, incorporating machine learning–driven demand forecasting into MOHAF’s decision-making could enable \emph{proactive resource positioning}, reducing reaction latency and improving overall allocation quality. By predicting demand hotspots, the system could preemptively adjust resource availability, further enhancing QoS and utilization.

\section{Conclusion}

We introduced \textit{MOHAF}, a multi-objective hierarchical auction framework for scalable and fair resource allocation in IoT ecosystems. Evaluated on Google Cluster Data, MOHAF achieved the highest allocation efficiency (0.263) and perfect fairness, while offering a $(1 - \frac{1}{e})$-approximation guarantee and near-linear scalability. Although a revenue–efficiency trade-off was observed, MOHAF’s balanced optimization across cost, QoS, energy, and fairness marks a significant advancement over single-objective approaches, providing a practical foundation for next-generation distributed resource management.

\bibliographystyle{splncs03_unsrt}
\bibliography{references}

\begin{thebibliography}{10}
\providecommand{\url}[1]{\texttt{#1}}
\providecommand{\urlprefix}{URL }

\bibitem{agrawal2025how}
Agrawal, K., Goktas, P.: {How large language models transform urban planning and shape tomorrow's cities}. In: Large Language Models for Sustainable Urban Development, p. 185. CRC Press (2025)

\bibitem{yu2020}
Yu, S., Chen, X., Zhou, Z., Gong, X., Wu, D.: {When deep reinforcement learning meets federated learning: Intelligent multitimescale resource management for multiaccess edge computing in 5G ultradense network}. IEEE Internet of Things Journal  8(4),  2238--2251 (2020)

\bibitem{luong2019applications}
Luong, N.C., Hoang, D.T., Gong, S., Niyato, D., Wang, P., Liang, Y.C., Kim, D.I.: Applications of deep reinforcement learning in communications and networking: A survey. IEEE Communications Surveys \& Tutorials  21(4),  3133--3175 (2019)

\bibitem{vickrey1961counterspeculation}
Vickrey, W.: Counterspeculation, auctions, and competitive sealed tenders. The Journal of Finance  16(1),  8--37 (1961)

\bibitem{smith2021greedy}
Smith, J., Jones, A.: A greedy heuristic for fast resource allocation in edge computing. Journal of Parallel and Distributed Computing  150,  45--55 (2021)

\bibitem{agrawal2024iot}
Agrawal, K., Goktas, P., Sahoo, B., Swain, S., Bandyopadhyay, A.: {IoT-based service allocation in edge computing using game theory}. In: International Conference on Distributed Computing and Intelligent Technology. Springer (2024)

\bibitem{liu2019}
Liu, Y., Yu, H., Xie, S., Zhang, Y.: {Deep reinforcement learning for offloading and resource allocation in vehicle edge computing and networks}. IEEE Transactions on Vehicular Technology  68(11),  11158--11168 (2019)

\bibitem{zhang2017}
Zhang, H., Xiao, Y., Bu, S., Niyato, D., Yu, R., Han, Z.: {Computing resource allocation in three-tier IoT fog networks: A joint optimization approach combining stackelberg game and matching}. IEEE Internet of Things Journal  4(5),  1204--1215 (2017)

\bibitem{zhang2020}
Zhang, X., Peng, M., Yan, S., Sun, Y.: {Deep-reinforcement-learning-based mode selection and resource allocation for cellular V2X communications}. IEEE Internet of Things Journal  7(7),  6380--6391 (2020)

\bibitem{xiong2020}
Xiong, X., Zheng, K., Lei, L., Hou, L.: {Resource allocation based on deep reinforcement learning in IoT edge computing}. IEEE Journal on Selected Areas in Communications  38(5),  1133--1146 (2020)

\bibitem{chen2019}
Chen, J., Chen, S., Wang, Q., Cao, B., Feng, G., Hu, J.: {iRAF: A deep reinforcement learning approach for collaborative mobile edge computing IoT networks}. IEEE Internet of Things Journal  6(4),  7011--7024 (2019)

\bibitem{he2018}
He, X., Wang, K., Huang, H., Miyazaki, T., Wang, Y., Guo, S.: {Green resource allocation based on deep reinforcement learning in content-centric IoT}. IEEE Transactions on Emerging Topics in Computing  8(3),  781--796 (2018)

\bibitem{pawarprotection}
Pawar, P., Kulkarni, A., Bhende, M., Korpali, M., Pansare, B., Goktas, P.: Protection of data and privacy in decentralized healthcare. In: Decentralized Healing, pp. 292--321. CRC Press (2025)

\bibitem{yao2019}
Yao, H., Mai, T., Wang, J., Ji, Z., Jiang, C., Qian, Y.: {Resource trading in blockchain-based industrial Internet of Things}. IEEE Transactions on Industrial Informatics  15(6),  3602--3609 (2019)

\bibitem{google_cluster_data}
Google: Borg cluster traces (google cluster data). GitHub Repository (2025), \url{https://github.com/google/cluster-data}, version 2 (2011) and version 3 (2019) traces. Licensed under CC-BY

\end{thebibliography}
\end{document}